\documentstyle[12pt,axodraw]{article}

\newcommand{\pr}[3]{{{\it Phys.~Rep.}~{#1} (#3) #2}}
\newcommand{\prl}[3]{{{\it Phys.~Rev.~Lett}~{#1} (#3) #2}}
\newcommand{\plb}[3]{{{\it Phys.~Lett.}~{\bf B#1} (#3) #2}}
\newcommand{\npb}[3]{{{\it Nucl.~Phys.}~{\bf B#1} (#3) #2}}
\newcommand{\prd}[3]{{{\it Phys.~Rev.}~{\bf D#1} (#3) #2}}
\newcommand{\ptp}[3]{{{\it Prog.~Theor.~Phys.}~{\bf #1} (#3) #2}}
\newcommand{\rmp}[3]{{{\it Rev.~Mod.~Phys.}~{\bf #1} (#3) #2}}
\newcommand{\df}[1]{{\rm d}#1}

\newcommand{\leqsim}{\,\raisebox{-0.6ex}{$\buildrel < \over \sim$}\,}
\newcommand{\geqsim}{\,\raisebox{-0.6ex}{$\buildrel > \over \sim$}\,}

\newcommand{\ba}{\begin{eqnarray}}
\newcommand{\be}{\begin{equation}}
\newcommand{\ea}{\end{eqnarray}}
\newcommand{\ee}{\end{equation}}

\newcommand{\etal} {\mbox{\em et al}}

\newcommand{\tev}{{\mbox{TeV}}}
\newcommand{\gev}{{\mbox{GeV}}}
\newcommand{\mev}{{\mbox{MeV}}}

\newcommand{\hc}{\mbox{h.c.}}

\begin{document}
\thispagestyle{empty}
\begin{flushright}
{\tt ULB-TH-96/21\\December 1996}
\end{flushright}
\vspace{15mm}
\begin{center}
{\LARGE $CP$ in $R$-parity violating models} \\ \vspace{15mm}
{\Large S.~A.~Abel}\\
\vspace{1cm}
{\small\it Service de Physique Th\'eorique, 
Universit\'e Libre de Bruxelles\\
Boulevard du Triomphe, Bruxelles 1050, Belgium }\\

\end{center}
\vspace{2cm}
\begin{abstract}
\noindent 
It is shown that in $R$-parity violating models of 
supersymmetry, the CP violation observed in the kaon 
system could arise purely from the $R$-parity violating 
scalar interactions ($A$-terms), with no CP violation in the 
CKM matrix. The direct CP 
violating parameter, $\varepsilon'$, could be as large
or larger than that expected in the Standard Model.
CP violation in the $B$ system however is expected to be negligible. 
\end{abstract}
\vspace{4cm}


\newpage

\section{Introduction}

In the Standard Model (SM)  the only possible source for 
the CP violation observed in the kaon system is in the 
Yukawa couplings. By a suitable redefinition of the 
fields, the CP violation can be described by a single phase 
appearing in the charged-current couplings ({\em i.e.} in the 
Kobayashi-Maskawa matrix). The fact that this phase can exist only 
if there are three (or more) generations is often seen as 
providing some support for the KM mechanism~\cite{km}. 

In supersymmetry however, there are many other possible 
sources of CP violation, particularly in the scalar 
couplings~\cite{fg}.  
Despite this, the usual (sometimes referred to as `minimal')
choice of degenerate and real soft-supersymmetry breaking 
terms neglects these, and simply incorporates the KM mechanism. 
I shall refer to this model as the `Constrained' Minimal 
Supersymmetric Standard Model (CMSSM). 

There are two reasons why the CMSSM receives the lion's share of 
attention in work on supersymmetry (apart from the fact that it has
the fewest parameters). Firstly it arises in the simplest supergravity 
models in which the supersymmetry breaking occurs in a hidden sector 
and is transmitted gravitationally to the visible sector. Secondly, the 
realness and degeneracy of the supersymmetry breaking couplings 
protects against large FCNCs and electric dipole moments (EDMs). 

The first of these reasons is probably unjustified, since in the 
light of string theory, it has become apparent that there are 
many other patterns of supersymmetry breaking which can 
occur~\cite{christoph,bailin}. 
The second reason has of course the drawback that the CMSSM restricts 
(by construction) significant supersymmetric contributions to 
CP violation in the kaon system (typically to a fraction of those observed, 
and those expected in the SM)~\cite{goto}. 

In the CMSSM therefore, one is effectively {\em forced} to use the KM 
mechanism to explain the CP violation observed for kaons. In 
fact this is true 
even when one keeps the degeneracy but allows a CP violating 
phase in the supersymmetry breaking terms~\cite{mejm}.

In a recent paper, an alternative pattern of supersymmetry breaking  
was proposed for the MSSM,
in which the observed CP violation {\em can} arise only from 
the scalar couplings~\cite{mejm}. This relied on the fact that 
EDMs provide rather restrictive
constraints on the generation diagonal scalar couplings, but that 
the CP violation observed in the kaon system is a flavour-changing 
process. Here FCNCs provide relatively weak constraints on the generation 
off-diagonal scalar couplings. So with a judicious choice of supersymmetry
breaking terms, it is possible to generate $\varepsilon $ through supersymmetric 
box diagrams, without generating large EDMs. It was also argued that this 
type of off-diagonal texture for the scalar couplings should 
be justifiable in string effective 
supergravity simply by a choice of quantum numbers~\cite{christoph,bailin}.  

This paper considers another mechanism for generating $\varepsilon $ 
from scalars, in the context of $R$-parity violating models. 
Because they introduce new interactions, these models lead to 
flavour changing scalar-exchange diagrams, which can produce the observed
$\varepsilon $. As a picture of CP violation, this is more akin 
to the original `superweak' models
in which a $\Delta S=2 $ boson was exchanged~\cite{wolfi,cheng}. 
Here the boson in question 
is the sneutrino. The experimental signature can be different 
to that in the model of Ref.\cite{mejm}; certain choices of 
$R$-parity violating couplings can give measurable contributions to 
$\varepsilon '$, but the CP violation in the $B$-system is again expected 
to be insignificant. 

There are good reasons why explaining CP violation 
by scalar interactions is attractive. 
CP is a symmetry of pure gauge theory + fermions. 
In gauge theory, CP violation is possible only 
on the breaking of gauge symmetry by some scalars (composite 
or otherwise), when Yukawa couplings give masses to fermions. 
CP is also thought to be a discrete gauge symmetry of string theory. 
Hence CP violation does not occur in the Yukawa couplings of 
string theory at lowest order, and if it is 
to appear anywhere it seems that it must do so via spontaneous CP 
violation~\cite{dineetal}. If CP is spontaneously 
broken by moduli, this can naturally show up in the (supersymmetry 
breaking) scalar interactions~\cite{bailin}. 
Finally CP violation through the KM mechanism does not lead to a 
natural explanation of the only other observed CP violating phenomenon, 
the existence of a baryonic rather than anti-baryonic Universe. 

The next section introduces the $R$-parity violating model, 
and describes the restrictions on the new Yukawa couplings. Following 
this the new CP-violating diagrams contributing to 
the observed value of $\varepsilon $ will be considered. Finally the 
possible contributions to $\varepsilon '$ will be determined, and it will be 
shown that $\varepsilon '$ can be measurable in certain cases. 

\section{$R$-parity violation}

$R$-parity was introduced into supersymmetry to prevent
fast proton decay, and may be described as the invariance 
of the lagrangian derived from the MSSM superpotential  
\be
\label{superyuk}
W_{yuk}=\epsilon \left(\lambda_E H_1 L E^c
+ \lambda_D H_1 Q D^c
+ \lambda_U Q H_2 U^c +\mu H_1 H_2 \right) 
\ee
where colour and generation indices have been suppressed, under the 
transformation on the fields, $f_i$,
\be
f_i \rightarrow (-1)^{(2 S_i + 3 B_i + L_i)} f_i 
\ee 
where $f_i$ is an arbitrary field of spin $S_i$, baryon number
$B_i$ and lepton number $L_i$. (Here $\epsilon $ is a Levi-Cevita
symbol, and the superfields are defined as 
$Q(3,2,\frac{1}{6})$, 
$U^c(\overline{3},1,-\frac{2}{3})$, 
$D^c(\overline{3},1,\frac{1}{3})$, 
$L(1,2,-\frac{1}{2})$, 
$E^c(1,1,1)$, 
$H_1(1,2,-\frac{1}{2})$ and 
$H_2(1,2,\frac{1}{2})$, so that the conventional fermion 
mass matrices are given by $m_U = \lambda_U^* \langle h_2^0 \rangle $,
 $m_D = \lambda_D^* \langle h_1^0 \rangle $,
 $m_E = \lambda_E^* \langle h_1^0 \rangle $.)
 
This symmetry is often imposed 
on the lagrangian in order to forbid other baryon number 
violating operators which would lead to proton decay. 
The net effect of $R$-parity, is to allow a single superpartner to decay to 
only an odd number of lighter superpartners plus any number of SM particles.
Thus the lightest supersymmetric particle (LSP) will be absolutely stable, and 
provides a good candidate for the cold component of dark matter.
The other operators which are allowed in the superpotential by gauge 
symmetry, but are
forbidden by $R$-parity are 
\be 
U^c D^c D^c \mbox{ ; }\epsilon L L E^c \mbox{ ; } \epsilon L Q D^c
\mbox{ ; } \mu_i \epsilon L_i H_2  ,
\ee 
where in the first term there is an implicit Levi-Cevita symbol 
summing over colour. 
However not all of them need to be forbidden in order to 
prevent proton decay, and in fact one can safely add either
\be
W_L = \epsilon \left( \frac{1}{2} \lambda_{ijk} L_{i} L_{j} E^c_{k}
+ \lambda_{ijk} ' L_{i} Q_{j} D_{k}^c \right) 
\ee
or
\be
W_B = \frac{1}{2} \lambda_{ijk} '' U^c_{i} U^c_{j} D^c_{k},
\ee
where $ijk$ are generation indices, to the MSSM superpotential.
From now on we take $\mu_i = 0$ to avoid large masses for the heaviest 
neutrino (generically $1\mev\leqsim \nu_\tau\leqsim 1\gev$). 
Generally this implies that $(\mu,\mu_i)$ and 
$\langle (h^0_1,\tilde{\nu}_{Li})\rangle $ should be sufficiently 
aligned and, as discussed in Ref.\cite{nardi}, this may be ensured 
using horizontal symmetries. The following analysis is not further
constrained by, and is independent of the considerations in 
Ref.\cite{nardi}.

The (equally valid) terms in $W_L$ and $W_B$
allow baryon violation or lepton violation but not both. They are said to be 
$R$--parity
breaking~\cite{rpv,nardi,peter,ibanross,herb,herbi2,herbi3}. 
The first is particularly interesting 
and corresponds to the lagrangian being invariant under the 
following transformation known as {\em baryon--parity} \cite{ibanross,herb},
\be
f_i \rightarrow (-1)^{(2 S_i + 3 B_i)} f_i .
\ee
We shall concentrate on this case in which the superpotential looks like 
\ba
W &=& 
\epsilon \left(
\lambda_E H_1 L E^c 
+\lambda_D H_1 Q D^c
+\lambda_U Q H_2 U^c  
\right.
\nonumber\\
&& \left.+\frac{1}{2} \lambda L L E^c
 + \lambda ' L Q D^c
 + \mu H_1 H_2 \right).
\ea
The reason for this preference is that the generation 
of $\varepsilon $ will rely on the exchange of sneutrinos. Because they
do not couple strongly, sneutrinos tend to be much lighter than squarks
at the weak scale (given the assumption of nearly degenerate scalar masses
at the GUT scale).

The soft supersymmetry breaking terms appearing in the lagrangian are 
as follows
\begin{eqnarray}                                               
\label{susybreak}
-\delta {\cal{L}}&=&m^2_{ij}\phi_i \phi_j^*  + \frac{1}{2}M_A
\lambda_A \lambda_A \nonumber\\ 
&& +\epsilon \left( A_E h_1 {\tilde l} {\tilde e}^c 
   + A_D  h_1 {\tilde q} {\tilde d}^c
   + A_U  {\tilde q} h_2 {\tilde u}^c 
   + B \mu h_1 h_2 \right. \nonumber\\
&& + \left. \frac{1}{2} C {\tilde l} 
        {\tilde l} {\tilde e}^c
   + C'{\tilde l} {\tilde q} {\tilde d}^c
   +\hc \right)
\end{eqnarray}    
The $A$ and $C$ terms are trilinear scalar couplings, and the $m^2_{ij}$ and $M_A$ are 
masses for the scalars (generically denoted by $\phi_i$) and gauginos
respectively.

The coupling of interest here is $\lambda'$ since this is the one which 
mediates sneutrino exchange between down quarks. 
So with this choice of superpotential, the new terms in the 
lagrangian which are important from the point of view of 
CP violation in the $K$ system, are the following; 
\ba
\label{newterms}
- \delta {\cal L} &=& 
\lambda '_{ijk} \left(
  \overline{\nu}^c_{Li} d_{Lj} \tilde{d}^*_{Rk} 
- \overline{e}^c_{Ri}   u_{Lj} \tilde{d}^*_{Rk} +
  \nu_{Li} \tilde{d}_{Lj} \overline{d}_{Rk} 
- e_{Li}   \tilde{u}_{Lj} \overline{d}_{Rk} +
  \tilde{\nu}_{Li} d_{Lj} \overline{d}_{Rk} 
- \tilde{e}_{Li}   u_{Lj} \overline{d}_{Rk} \right.\nonumber\\
 && \left. + A'_{ijk} \tilde{\nu}_{Li} \tilde{d}_{Lj} \tilde{d}^*_{Rk} 
    - A'_{ijk} \tilde{e}  _{Li} \tilde{u}_{Lj} \tilde{d}^*_{Rk} \right) + \hc + \ldots 
\ea
For the sake of argument, the trilinear 
$C$-terms have been defined to include the Yukawa coupling.
There are various constraints on the new Yukawa couplings coming from 
both direct and indirect ({\em i.e.} renormalisation group effects, 
and contributions to the mass matrices) 
sources. Important bounds on $\lambda'$ have been derived from the 
contribution of the new couplings to the sneutrino mass-squared terms, via 
the renormalisation group running of the couplings. These contributions may 
push the sneutrino mass below its current experimental bound.
This was discussed in Ref.\cite{peter} assuming degenerate supersymmetry 
breaking, of the form 
\ba
\label{degen}
A_{U_{ij}}&=&A \lambda_{U_{ij}}\nonumber\\
A_{D_{ij}}&=&A \lambda_{D_{ij}}\nonumber\\
A_{E_{ij}}&=&A \lambda_{E_{ij}}\nonumber\\
C'_{ijk}        &=&A'\lambda'_{ijk}\nonumber\\ 
m^2_{ij}   &=&\delta_{ij} m_0^2\nonumber\\
M_A      &=&m_{1/2},
\ea
where $A$, $A'$, $B$, $m_{1/2}$ and $m_0$ have dimensions of mass and are
all less than $\sim$ 1~TeV in order to protect the weak scale.
To prevent large EDMs for the neutron, $A$, $B$ and $m_{1/2}$ are 
taken to be real. Later more general $A'$ couplings will be 
required, but the other supersymmetry breaking terms 
will remain as they are here.
The corresponding low energy values of the sneutrino masses 
may be approximated by running the RGEs numerically~\cite{peter},
\be
m^2_{L_i} = \frac{1}{2}M_Z^2\cos2\beta + m_0^2 + 0.51 m_{1/2}^2
 - \lambda^{\prime 2}_{ijk}(M_{GUT})
 \left( 13m_0^2 + 49m_{1/2}^2 - 1.5 m_{1/2}A' - 12 A'^2  \right) .
\ee
In Ref.\cite{peter} it was found that 
\be
\lambda'_{ijk} (M_{GUT} )\leqsim 0.15
\ee
is enough to prevent the sneutrino masses becoming too light. The Yukawa 
couplings are enhanced by a factor of 3.5 by the time they reach the 
weak scale, so that we shall adopt the bound
\be
\label{bound}
\lambda'_{ijk}(M_Z)\leqsim 0.5,
\ee
for all indices $i~j~k$. In addition there is the danger that renormalisation 
effects will induce an $m^2_{L_i H_1}$ term and cause the sneutrinos to 
get a VEV. Such a VEV can cause all manner of problems. \\
\begin{picture}(375,250)(0,40)
\Line(120,150)(320,150)
\DashCArc(220,150)(60,0,180)5
\Photon(145,225)(180,195)44
\Text(220,140)[]{ $\tilde{g}$}
\Text(220,225)[]{ $C'_{i11}\langle \tilde{\nu}_i\rangle$}
\Text(180,180)[]{ $\tilde{d}_L$}
\Text(260,180)[]{ $\tilde{d}_R$}
\Text(130,160)[]{ $d_L$}
\Text(310,160)[]{ $d_R$}
\Text(220,212)[]{ $\times$}
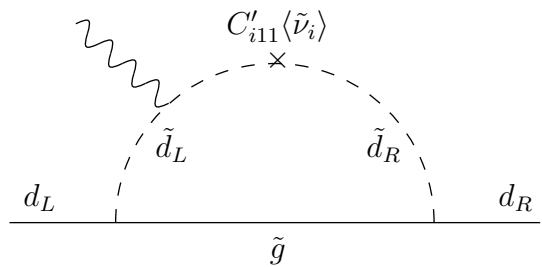
\Text(220,75)[]{\em figure (1): Possible contribution to neutron EDM from sneutrino VEV.}
\end{picture}\\
\noindent
For example, the 
the diagram shown in figure (1) can lead to large imaginary 
contributions to the neutron 
EDM.
Imposing 
\be
\label{bound2}
\lambda'_{i33} \approx 0,
\ee
is sufficient to prevent this~\cite{peter}. 
These bounds of course depend on the values of $m_0$ and in most cases may 
be relaxed somewhat. Quite restrictive bounds also come from the 
(non-observation) of the rare decay $\mu\rightarrow e\gamma $; 
for example~\cite{peter} 
\be
\label{bound3}
\lambda '_{1jk}(M_Z)\lambda '_{2jk}(M_Z) \leqsim 5 \times 10^{-4}
\left( \frac{\tilde{m}}{100 \gev }\right)^2 \mbox{  \hspace{1cm}} j,k=1,2
\ee
where $\tilde{m}$ is a measure of the squark masses.  
If all the $\lambda'_{ijk}$ couplings other than $\lambda'_{i33}$
are the same, then
for squark/gluino masses less than $\sim 1.4\tev$ this bound  
is the most restrictive. Other bounds may also 
been derived from $b\rightarrow s\gamma $, the most important of which 
is~\cite{peter}, 
\be
\lambda '_{ij2}(M_Z)\lambda '_{ij3}(M_Z) \leqsim 0.003
\left( \frac{\tilde{m}}{100 \gev }\right)^2.
\ee
For this analysis the relevant  products will be 
$\lambda'_{i12} \lambda'_{i21}$ and $\lambda'_{i11} \lambda'_{i21}$.
The only experimental limits on these come from the contribution to 
$\Delta m_K$ itself.  

Hopefully it will be possible in the near future to probe many of these 
couplings (for example $\lambda'_{1j1} $) individually at HERA,
if the squark masses are less than 300\gev~\cite{herbi2}. 
There are also bounds from cosmological considerations (see Ref.\cite{herb}
and references therein). However, these are strongly dependent on 
the assumed scenario of baryogenesis, and so will not be employed.
(In fact for one baryogenesis scheme which has recently been suggested, 
the couplings are not constrained by this at all~\cite{sarkar}.)

\section{CP violating parameters in the kaon system} 

Let us go on to consider the new contributions to $\Delta m_K$ and 
$\varepsilon $. These $\Delta S= 2 $ processes 
receive contribution from the sneutrino exchange diagrams 
shown in figure (2). By
assumption all the Yukawa couplings have no imaginary part
and so the tree level diagram contributes only to $\Delta m_K$.\\
\begin{picture}(430,250)(0,40)
\ArrowLine(10,130)(70,130)
\ArrowLine(70,220)(10,220)
\ArrowLine(70,130)(130,130)
\ArrowLine(130,220)(70,220)
\DashLine(70,130)(70,220)5
\Text(65,175)[]{$\scriptstyle \tilde{\nu}_i $}
\Text(15,136)[]{$\scriptstyle d_L$}
\Text(130,136)[]{$\scriptstyle s_R$}
\Text(15,226)[]{$\scriptstyle s_L$}
\Text(130,226)[]{$\scriptstyle d_R$}
\ArrowLine(160,130)(190,130)
\ArrowLine(220,220)(160,220)
\ArrowLine(250,130)(280,130)
\ArrowLine(280,220)(220,220)
\Line(190,130)(250,130)
\DashCArc(220,130)(35,0,180)5
\DashLine(220,165)(220,220)5
\Text(215,190)[]{$\scriptstyle \tilde{\nu}_i $}
\Text(220,125)[]{$\scriptstyle \tilde{g} $}
\Text(183,152)[]{$\scriptstyle \tilde{d}_L$}
\Text(257,152)[]{$\scriptstyle \tilde{s}_R$}
\Text(165,136)[]{$\scriptstyle d_L$}
\Text(280,136)[]{$\scriptstyle s_R$}
\Text(165,226)[]{$\scriptstyle s_L$}
\Text(280,226)[]{$\scriptstyle d_R$}
\ArrowLine(310,130)(370,130)
\ArrowLine(340,220)(310,220)
\ArrowLine(370,130)(430,130)
\ArrowLine(430,220)(400,220)
\Line(340,220)(400,220)
\DashCArc(370,220)(35,180,360)5
\DashLine(370,130)(370,185)5
\Text(365,160)[]{$\scriptstyle \tilde{\nu}_i $}
\Text(370,227)[]{$\scriptstyle \tilde{g} $}
\Text(338,195)[]{$\scriptstyle \tilde{s}_L$}
\Text(404,195)[]{$\scriptstyle \tilde{d}_R$}
\Text(315,136)[]{$\scriptstyle d_L$}
\Text(430,136)[]{$\scriptstyle s_R$}
\Text(315,226)[]{$\scriptstyle s_L$}
\Text(430,226)[]{$\scriptstyle d_R$}
\Text(220,75)[]{\em figure (2): Contibutions to $\Delta S=2$ processes.}
\end{picture}\\
\noindent
If CP violation arises only in the trilinear couplings 
(specifically in $A'_{ijk}$), then it is clear that the one loop diagrams 
must be responsible for $\varepsilon $. 
The one loop gluino contribution to the $\lambda'_{ijk} \tilde{\nu}_{Li} 
d_{L_j}\overline{d}_{R_k}$ coupling is found to be 
\be
\delta \lambda'_{ijk}=\lambda'_{ijk} F_1(x_{L_j}, x_{R_k} ) \frac{2\alpha_s}{3\pi}
\frac{A'_{ijk}}{m_{\tilde{g}}} 
\ee
where $x_i=\tilde{m}^2_i/m_{\tilde{g}}^2 $, and where $F_1$ is 
given by the integral,
\be
 F_1(x_j, x_k ) = \int^1_0\int^1_0  \frac{y \df y\df z }{(1-y)+y((1-z)x_i + z x_j)} .
\ee
In calculating this, the small mixing between the left and right down squarks 
has been neglected for simplicity. If for example $ x_{L_j} = x_{R_k} = x = 
\tilde{m}^2/m_{\tilde{g}}^2 $ then 
\be
F_1(x,x)= \frac{x-1-\log x}{(x-1)^2},
\ee
and for $x\leq 0.5 $ we have $F_1 \geqsim 1 $.
The diagrams in figure (2) lead to the operator 
\be
\delta {\cal H}_{eff}=\sum_i \frac{(\lambda'_{i21}+\delta \lambda'_{i21})
(\lambda'_{i12}+\delta \lambda_{i12}^{\prime *})}
{m^2_{\tilde{\nu}_i}} Q_4
\ee
where
\be 
Q_4=\overline{d}_R^\alpha
s_L^\alpha
\overline{d}_L^\beta
s_R^\beta.
\ee
The summation above is over the three generations of
exchanged sneutrinos, and $\alpha $ and $\beta $ are colour indices.
In order to simplify things, it will be assumed that one of the 
sneutrinos, $\tilde{\nu}_i $, couples more strongly through 
the $\lambda'_{ijk}$, and the summation will be dropped. 

Since there are many unknowns in the model, including the 
relative importance of the various sneutrino exchanges, 
it is only worth obtaining order of magnitude estimates here, and 
for the rest of the present discussion, the vacuum saturation 
approximation will be sufficient. Despite these
uncertainties, an order of magnitude estimate makes it possible to see
roughly how close to experimental bounds, or otherwise, the new couplings
must be if they are to explain the observed CP violation in the Kaon
system. We shall find that the direct CP violating parameter can be 
observable within the current experimental bounds which is, in itself, a
useful result. QCD effects are expected to contribute a factor of order a
few to the final estimate as in other discussions of FCNCs in 
supersymmetry. Since we do not anticipate any cancellations, the 
precise value of the $B$-factors should not be as crucial as in
the SM, and also the effects of operator mixing should not change 
our estimate significantly. 

Using the matrix element~\cite{gabbiani},
\be
\langle \overline{K} | Q_4 | K \rangle =  \frac{m_K f_K^2}{m^2_{\tilde{\nu}_i}}
\left( 
\frac{1}{24}+\frac{1}{4}\left(\frac{m_K}{m_s+m_d}\right)^2
\right),
\ee
where the vacuum insertion approximation has been made,
the $\Delta S=2 $ parameters are found to be,
\ba
\Delta m^{SUSY}_K &=& 2 \mbox{Re}(M_{12}) \nonumber\\
&=& \lambda'_{i21}\lambda'_{i12} 
\frac{m_K f_K^2}{m^2_{\tilde{\nu}_i}}
\left( 
\frac{1}{12}+\frac{1}{2}\left(\frac{m_K}{m_s+m_d}\right)^2
\right),
\ea
and 
\ba
\varepsilon  &=& - \frac{e^{i\pi/4}}{2\sqrt{2}}
\frac{\mbox{Im}(M_{12})}{\mbox{Re}(M_{12})}\nonumber\\
&=& - \frac{e^{i\pi/4} \alpha_s F_1(x_R,x_L)}
{3\pi \sqrt{2}}
\frac{\mbox{Im}(A'_{i21}-A'_{i12})}
{m_{\tilde{g}}} \left(\frac{\Delta m^{SUSY}_K}{\Delta m_K}\right).
\ea
Assuming degeneracy in $x_{L_i}$ and $x_{R_i}$ leads to errors 
of at most ${\cal{O}}(m_b^2/m_{\tilde{g}^2})$. 
As in the model of CP proposed in Ref.\cite{mejm}, the generation 
of $\varepsilon $ relies on having a sufficient off-diagonality in the 
trilinear terms.
Since the prefactor in $\varepsilon $ is $\sim 10^{-2}$, this implies
that some non-degeneracy and asymmetry is required in the 
$A'_{ijk}$ matrices at the 
GUT scale (the RGEs being unlikely to generate enough). 
It is also clear that the supersymmetric contribution to $\Delta m_K$ must 
be substantial. This last point ($\Delta m_K^{SUSY} \approx \Delta m_K$)
leads to an estimate of a product of $R$-parity violating couplings, 
\be
\label{estimate}
\lambda'_{i12} 
\lambda'_{i21} \approx 10^{-9} \left( \frac{m_{\tilde{\nu}_i}}
{100\gev}\right)^2,
\ee
which is in accord with the bounds coming from the same quantity 
in Refs.\cite{rpv,nardi,peter}. This is easily compatible with all 
known bounds on $R$-parity violating couplings.\\
\begin{picture}(430,250)(0,40)
\ArrowLine(210,150)(150,150)
\ArrowLine(140,220)(80,220)
\ArrowLine(150,150)(210,115)
\ArrowLine(200,220)(140,220)
\DashLine(150,150)(140,220)5
\Vertex(150,150)3
\Vertex(140,220)3
\Text(140,185)[]{$\scriptstyle \tilde{\nu}_i $}
\Text(210,123)[]{$\scriptstyle d_R$}
\Text(210,156)[]{$\scriptstyle d_L$}
\Text(85,226)[]{$\scriptstyle s_L$}
\Text(200,226)[]{$\scriptstyle d_R$}
\ArrowLine(380,150)(320,150)
\ArrowLine(310,220)(250,220)
\ArrowLine(320,150)(380,115)
\ArrowLine(370,220)(310,220)
\DashLine(320,150)(310,220)5
\Vertex(320,150)3
\Vertex(310,220)3
\Text(310,185)[]{$\scriptstyle \tilde{\nu}_i $}
\Text(380,123)[]{$\scriptstyle d_R$}
\Text(380,156)[]{$\scriptstyle d_L$}
\Text(255,226)[]{$\scriptstyle s_L$}
\Text(370,226)[]{$\scriptstyle d_R$}
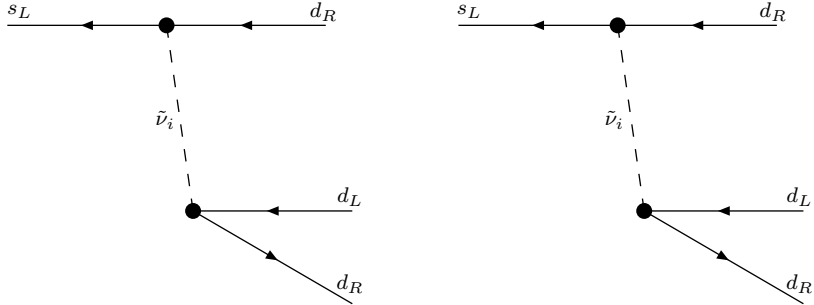
\Text(220,75)[]{\em figure (3): New contributions to $\Delta S=1$ processes; The vertices}
\Text(242,60)[]{\em include one-loop squark/gluino corrections.}
\end{picture}\\
\noindent
Now consider the $\Delta S=1 $ contributions to the effective 
hamiltonian. These occur through the  sneutrino exchange diagrams
shown in figure (3) and corresponding diagrams with up quarks and 
selectron exchange, which lead to operators of the form 
\be
Q= \overline{d}_R^\alpha
s_L^\alpha
\overline{d}_L^\beta
d_R^\beta.
\ee
in the effective potential. A Fierz rearrangement shows this to be 
similar to the operator $Q_8$ of Ref.\cite{buchalla}, and the 
isospin two part of the matrix element $\langle \pi \pi 
|{\cal{H}}_{eff}|K\rangle $) is therefore of order;
\be
f_\pi m_K^2 \left(
\sqrt{3} \left(\frac{m^2_K}{m_s + m_d}\right)^2 - \frac{1}{2\sqrt{3}}
\left( 1-\frac{m_\pi^2}{M^2_K} \right) \right).
\ee
Again, since we are ignorant of the relative contributions of the
sneutrino and selectron exchanges, the vacuum saturation approximation
is sufficient to obtain an order of magnitude estimate. 
The parameter $\frac{\varepsilon '}{\varepsilon} $ is given by 
\ba
\left| \frac{\varepsilon'}{\varepsilon}\right|
&= & \frac{\omega}{\sqrt{2}|\varepsilon|}
\left|     \frac{\mbox{Im}(A_0)}{\mbox{Re}(A_0)}
-          \frac{\mbox{Im}(A_2)}{\mbox{Re}(A_2)}\right|
\nonumber\\
&\sim &
\frac{\sqrt{6} \Delta m_K }{\mbox{Re}(A_0)}
\frac{f_\pi m_K}{f^2_K}
\frac{\mbox{Im}(A'_{i21}+A^{\prime *}_{i11})}
     {\mbox{Im}(A'_{i21}+A^{\prime *}_{i12})}
\frac{\lambda'_{i11} \lambda'_{i21}}
     {\lambda'_{i12} \lambda'_{i21}}.
\ea
Since the contributions to the isospin-2 and isospin-0 
parts are comparable, there is no factor of $\omega=1/22$ in 
this expression since the isospin-2 contribution is dominant here
(provided that long distance contributions do not contribute too much 
to the relative sizes of Re$(A_0)$ and Re$(A_2)$). 
There is no credible mechanism whereby the different $A'_{ijk}$ can 
differ by the four orders of magnitude which will be required if 
$\varepsilon '$ is to be significant. Instead there must be a 
hierarchy in the Yukawa couplings for this to be the case.
For example, setting $A'_{i11}=A'_{i12}$,
and again taking exchanges in the $i$'th generation 
to be dominant, one finds an order of magnitude estimate for 
$\varepsilon' $; 
\be
\left| \frac{\varepsilon'}{\varepsilon}\right|
 \sim  10^{-7} \frac{\lambda '_{i11}}{\lambda'_{i12}}
\ee
where the value Re$(A_0)=2.7\times 10^{-7}\gev $ has been used.
Together with the estimate in Eq.(\ref{estimate}), 
this means that for the product $\lambda'_{i11} 
\lambda'_{i21} $,
\be
\lambda'_{i11} 
\lambda'_{i21} \sim 0.01 \left| \frac{\varepsilon'}{\varepsilon}\right|
\left( \frac{m_{\tilde{\nu}_i}}
{100\gev}\right)^{-2}.
\ee
Clearly this is compatible with both `superweak' scenarios 
( $\frac{\varepsilon'}{\varepsilon}=0 $) and with `miliweak' 
theories such as the Standard Model which predicts 
$| \frac{\varepsilon'}{\varepsilon}|= few\times 10^{-4}$.
It is even consistent with values of $| \frac{\varepsilon'}{\varepsilon}| $
{\em larger} than that predicted in the Standard Model. This is
of interest in the light of the high values for $\varepsilon '$ 
currently claimed by the NA31 collaboration.

\section{Discussion}

What do these estimates mean for the new $R$-parity violating couplings? 
Clearly little can be said about the $\lambda_{ijk}$ couplings. 
For the $\lambda'_{ijk}$ couplings, the values for $\varepsilon $ and 
$\varepsilon' $ together with the bound coming from the sneutrino 
masses in Eq.(\ref{bound}) and VEVs in Eq.(\ref{bound2}) imply  
that there must be some hierarchy 
in the couplings if $\varepsilon' $ is to be measurable
(although this is as nothing compared to the hierarchy 
in $\lambda _U$, $\lambda_D $ and $\lambda_E $). 
Interestingly however, there already exist in the literature examples where 
such a hierarchy is partially generated. In Ref.\cite{herbi3}, it was 
found that in models of spontaneously broken gauged $R$-symmetry, the 
low energy models break $R$-parity. Furthermore, the discipline of 
anomaly cancellation forbids $\lambda'_{ijk}$ couplings involving the 
third generation. The generation of a larger $\varepsilon '$ in these 
models would therefore be a natural possibility. 

The picture of CP violation  
described in this paper should be easily differentiable
from the CMSSM and the SM. It is expected that there will be enough 
experimental data to considerably overdetermine the Kobayashi-Maskawa 
matrix. In addition, the contribution to CP violation 
in the $B$-system should again be small; in the Standard Model 
and the MSSM, this occurs at tree-level, whereas in this case 
CP violation may only be generated through one-loop diagrams. 

This paper, as Ref.\cite{mejm}, highlights 
the fact that the current 
`paradigm' for the MSSM is perhaps a little too constrained from the 
point of view of FCNCs and CP violating phenomena. In fact there is 
room to generate all observed CP violating phenomena from the scalar 
sector alone. Relaxing the assumption of $R$-parity allows the 
possibility of {\em direct} CP violation which can even be larger than 
that predicted in the Standard Model. 

It should be stressed therefore that the adoption of the 
Kobayashi-Maskawa model of CP violation is not the `minimal' 
choice for supersymmetry. It is no longer preferred 
from a theoretical point of view, and there is no definition of 
`naturalness' under which it is somehow more qualified. 
Furthermore siting the CP violating in the scalar sector
of supersymmetry, may lead to a better understanding of other
questions concerning CP violation such as baryogenesis, and 
the strong CP problem. 

Finally, given that the CP here is spontaneously broken, there is 
the hope of a solution to the strong CP problem. At the 
Planck scale the value of $\overline{\theta}$ is naturally zero, and
here it is clear that all radiative corrections are
suppressed. This question arises because of the obvious similarity 
of this model to the strong CP solutions discussed in ref.\cite{cheng}. 
There the strong CP problem was solved if the $\Delta S=2 $ 
boson maintained a very small VEV. The boson in this case is 
the sneutrino and it may be that its VEV remains small enough even in
models with broken $R$-parity. This will be the subject of future work. 

\bigskip
\vspace{1cm}

\noindent{\bf Acknowledgments}

\noindent I would like to thank H.~Dreiner, J.-M.~Fr\`ere and
P.~L.~White for their comments and suggestions.

\newpage

\end{document}